\documentclass{PoS}
\usepackage[utf8]{inputenc}
\usepackage[verbose]{placeins}
\usepackage{amsmath}
\usepackage{amssymb}
\usepackage{amsfonts,mathrsfs,fixmath,amsthm}
\usepackage{bbm}
\usepackage{color}
\usepackage[sort&compress]{natbib}
\bibpunct{[}{]}{,}{n}{}{}
\newcommand{\Dov}{\ensuremath{D_\text{ov}}}
\newcommand{\Dw}{\ensuremath{D_\text{w}}}

\newcommand{\ket}[1]{\ensuremath{\left|#1\right\rangle}}
\newcommand{\bra}[1]{\ensuremath{\left\langle#1\right|}}
\newcommand{\braket}[2]{\ensuremath{\left\langle\left.#1\right| #2
    \right\rangle}}

\newcommand{\RR}[1]{\ensuremath{\ket{R_{#1}}}}
\newcommand{\LL}[1]{\ensuremath{\bra{L_{#1}}}}
\newcommand{\dRR}[1]{\ensuremath{\ket{\pt R_{#1}}}}
\newcommand{\dLL}[1]{\ensuremath{\bra{\pt L_{#1}}}}
\newcommand{\pt}{\ensuremath{\partial_{t}}}

\newcommand{\bA}{\ensuremath{\mathcal{B}}}
\newcommand{\upperRomannumeral}[1]{\uppercase\expandafter{\romannumeral#1}}
\newcommand{\Tr}{\ensuremath{\operatorname{tr}}}
\newcommand{\CC}{\ensuremath{\mathbb{C}}}
\newcommand{\CNN}{\ensuremath{\mathbb{C}^{n \times n}}}

\newcommand{\MeV}{\ensuremath{\text{MeV}}}

\newcommand{\scsc}{\ensuremath{\sigma_{\text{\tiny CSE}}}}
\DeclareMathOperator{\sgn}{sgn}

\DeclareMathOperator{\diag}{diag}

\newcommand{\lr}[1]{\left(#1\right)}

\graphicspath{ {./plots_dir/} {./plots/} {./} }

\title{A method to compute derivatives of functions of large complex matrices}

\ShortTitle{A method to compute derivatives of functions of large complex matrices}

\author{\speaker{Matthias Puhr}
and Pavel Buividovich  \thanks{This work was supported by the S.~Kowalevskaja award from
the Alexander von Humboldt foundation.}\ \\
Institute of Theoretical Physics, Regensburg University, \\ 93040 Regensburg, Germany \\
\email{matthias.puhr@physik.uni-regensburg.de},\\ \email{pavel.buividovich@physik.uni-regensburg.de} }

\abstract{A recently developed numerical method for the calculation of derivatives of
functions of general complex matrices, which can also be combined with implicit matrix
function approximations such as Krylov-Ritz type algorithms, is presented. An important
use case for the method in the context of lattice gauge theory is the overlap Dirac
operator at finite quark chemical potential. Derivatives of the lattice Dirac operator are
necessary for the computation of conserved lattice currents or the fermionic force in
Hybrid Monte-Carlo and Langevin simulations. To calculate the overlap Dirac operator at
finite chemical potential the product of the sign function of a non-Hermitian matrix with
a vector has to be computed. For non-Hermitian matrices it is not possible to efficiently 
approximate the sign function with polynomials or rational functions. Implicit approximation
algorithms, like Krylov-Ritz methods, that depend on the source vector have to be used
instead. Our method can also provide derivatives of such implicit approximations.  We show
how a generalised deflation prescription can be used to improve the performance of the
method, if some eigenvalues and eigenvectors of the matrix being differentiated are
known. To show that the method is efficient and well suited for practical calculations we
provide test results for the two-sided Lanczos approximation of the finite-density overlap
Dirac operator on SU(3) gauge field configurations on lattices with sizes up to $14 \times
14^3$.}

\FullConference{34th annual International Symposium on Lattice Field Theory\\
         24-30 July 2016\\
         University of Southampton, UK}

\begin{document}

\section{Introduction}
Many problems in science and engineering can be formulated with the help of matrix valued
functions of matrices~\cite{Higham2008}. In the field of lattice gauge theories one
prominent example for an application of matrix functions is the so-called overlap Dirac
operator. The overlap operator respects the Ginsparg--Wilson equation and makes is
possible to define chiral symmetry on the lattice. At finite (quark) chemical potential
the massless overlap Dirac operator is given by~\cite{Bloch:06:1}
\begin{equation}
\label{eq:overlap}
 \Dov := \frac{1}{a}\left(\mathbbm{1}+\gamma_5 \sgn\left[ H(\mu)
\right] \right),
\end{equation} with the kernel operator $H(\mu) := \gamma_5 \Dw(\mu)$, $\Dw(\mu)$ is the
Wilson--Dirac operator at chemical potential $\mu$, $\sgn$ is the matrix sign function and
$a$ stands for the lattice spacing. Evaluating the matrix sign function exactly for large
matrices is not feasible and for simulations with reasonably large lattices one has to rely on
approximation methods to calculate the overlap operator. Efficient approximation methods
exist for both Hermitian and non-Hermitian kernel operators. For some applications in
lattice gauge theory it is necessary to compute the derivative of the lattice Dirac
operator. Examples are the computation of conserved lattice currents or the evaluation of
the Fermionic force in  HMC and Langevin simulations. For Hermitian kernels one can
usually construct a sufficiently good approximation of the overlap operator by using a polynomial
or rational function approximation of the sign function. If the kernel is a non-Hermitian
matrix on the other hand one generally has to use implicit approximation methods, like
Krylov-Ritz type algorithms. For implicit function approximations it is not always clear
how to take the derivative of the approximation. Finite difference methods often suffer from severe
round-off errors and a naive application of algorithmic differentiation can lead to
numerically unstable algorithms~\cite{Buividovich:14:3}. In this contribution we present
a numerical differentiation method for functions of general complex matrices and show how
a generalised deflation prescription can be used to greatly improve the efficiency of the
method. Our method works with any matrix function approximation algorithm and can also be
used together with implicit approximation algorithms.

\section{Numerical derivatives of matrix functions}
\label{sec:derivatives}
Matrix valued functions of a matrix can be defined in several equivalent ways. A very
general approach is to define a matrix function via the
Jordan decomposition of the matrix. Any complex matrix $A \in \CNN$ can be brought to the
Jordan normal form 
\begin{equation}
  \label{eq:A_Jordan}
  X^{-1} A X = J = \diag\left(J_1,J_2,\dots,J_k\right),
\end{equation}
where the right hand side is a block diagonal matrix and every block matrix $J_i$
corresponds to an eigenvalue $\lambda_i$ of $A$. The $J_i$ are called Jordan blocks and
are given by 
\begin{equation}
  \label{eq:Jordan_block}
  J_i = J_i(\lambda_i) = \begin{pmatrix}
    \lambda_i & 1         &   0     &  \cdots    & 0       \\
         0    & \lambda_i &   1     &  \ddots    &  \vdots \\
         0    & \ddots    & \ddots  &  \ddots    &  0      \\
      \vdots  & \ddots    & \ddots  &  \lambda_i &  1      \\
          0   & \cdots    & 0       &    0       & \lambda_i
  \end{pmatrix} \in \CC^{m_i \times m_i},
\end{equation}
with $m_1 + m_2 + \dots + m_k = n$. The Jordan matrix $J$ is unique modulo permutations of
the blocks, but the transformation matrix $X$ is not. 
If the Jordan normal form of the matrix is known on can define the matrix function
as~\cite{Higham2008,Golub1996}
\begin{equation}
  \label{eq:matrix_func}
   f(A) := X f(J) X^{-1} = X \diag(f(J_i)) X^{-1}.
\end{equation}
The function of a Jordan block matrix has the form
\begin{equation}
  \label{eq:Jordan_function}
  f(J_i) := \begin{pmatrix}
    f(\lambda_i) &  f'(\lambda_i)  &  \dots   & \frac{f^{(m_i-1)}(\lambda_i)}{(m_i-1)!}  \\
          0      & f(\lambda_i)    &  \ddots  & \vdots                                  \\
    \vdots       &    \ddots       &  \ddots  & f'(\lambda_i)                           \\
          0      &    \cdots       &  0        & f(\lambda_i)
  \end{pmatrix}.
\end{equation}
\begin{sloppypar}
If $A$ is a diagonalisable matrix the Jordan matrix $J$ becomes a diagonal matrix with the
eigenvalues $\lambda_i$ on the main diagonal and equation~\eqref{eq:matrix_func} reduces
to the so-called spectral form ${f(A) = X \diag(f(\lambda_1), f(\lambda_2), \dots,
f(\lambda_n)) X^{-1}}$.  While the definition \eqref{eq:Jordan_function} and the spectral
form are very convenient to study matrix functions analytically, they are not well suited
for numerical computations. In practice it is however seldom necessary to explicitly know
the function of a matrix $f(A)$. In most cases it is sufficient to compute the result
$\ket{y}=f(A)\ket{x}$ of the action of the matrix function on a source vector
$\ket{x}$. 
\end{sloppypar}
If the matrix  $A =A(t)$ depends on a parameter $t$ then the action of the derivative
of the matrix function with respect to $t$ is given 
by\footnote{We assume that the source vector $\ket{x}$ does not depend on $t$}
$\partial_t \ket{y} = \lr{\partial_t f\lr{A\lr{t}}} \ket{x} $.  
A theorem by R. Mathias~\cite{Mathias1996} states that the following equation holds under
some mild assumptions that are generally fulfilled in lattice QCD calculations:
\begin{equation}
  \label{eq:mathias_derivative}
  f\left(\bA\right) = \left(\begin{array}{cc}
        f(A(0)) &  \pt  f(A(t))  \\
        0 & f(A(0))
      \end{array}\right), \quad \bA(A) := \left(
        \begin{array}{cc}
          A(0) &  \pt A(t)  \\
          0   & A(0)
        \end{array}
      \right)
\end{equation}
A notable feature of equation~\eqref{eq:mathias_derivative} is that one does not need to
know the explicit form of the function $\pt f(A)$ to compute the derivative of
$f(A)$. The price to pay is that one has to compute the function $f$ for a block matrix of
twice the size of the original matrix $A$. Since $\bA$ is a sparse matrix this is usually
not a problem in practical calculations. 
The big advantage of using equation~\eqref{eq:mathias_derivative} is that the derivative
can be computed by applying any approximation algorithm to the left hand side of the
following equation:
\begin{equation}
  \label{eq:func_derivative}
  f(\bA) \begin{pmatrix} 0 \\ \ket{x} \end{pmatrix} = \begin{pmatrix}
    \pt f(A) \ket{x} \\ \ \ \  f(A) \ket{x}  \end{pmatrix}
\end{equation}
 An efficient method to compute the matrix sign function for non-Hermitian matrices is the
two-sided Lanczos (TSL) method~\cite{Bloch2010,Bloch:08:2,Bloch:07:1,Bloch:11:2} and 
in~\cite{Buividovich:14:3} we demonstrated that the TSL can be used in conjunction with 
 equation~\eqref{eq:func_derivative} to compute the derivative of the finite density
 overlap operator. 

\section{Deflation}
\label{sec:deflation}
The efficiency of approximation algorithms of matrix functions in general depends on the
spectrum of the matrix in question. Evaluating the matrix function is equivalent to
computing equation~\eqref{eq:Jordan_function}  for all Jordan blocks and it is
obvious that eigenvalues close to a pole or a branch cut of the function
$f$ can lead to problems in a numerical algorithm. The sign function has a discontinuity
along the imaginary axis and it has been shown that the approximations of the
matrix sign function can be greatly improved by a so called 
``deflation''~\cite{Bloch2010,Bloch:08:2,Bloch:07:1}. In the deflation procedure the
eigenvectors corresponding to eigenvalues close to zero are projected out from the source
vector $\ket{x}$ and the sign function is evaluated exactly on the subspace spanned by
these eigenvectors:
\begin{equation}
  \label{eq:defl}
  \sgn(A)\ket{x} = \underbrace{\sum\limits_{i=1}^{m}
\sgn(\lambda_i)\ket{R_i}\braket{L_i}{x}}_{\text{exact}} + \underbrace{\sgn(A)P_{m}^n
\ket{x}}_{\text{approximation}},
\end{equation}
where $\RR{i}$ and $\LL{i}$ are the right and left eigenvectors of $A$ corresponding to
the eigenvalue $\lambda_i$ and $P_{m}^n = \sum_{i=m+1}^{n}\ket{R_i}\bra{L_i}$ is a
projector. The standard deflation procedure~\eqref{eq:defl} relies on the spectral
decomposition of $A$ and works only for diagonalisable
matrices. The block matrix $\bA$ is in general not diagonalisable and in the following we
briefly describe how to generalise the deflation procedure to non-diagonalisable
matrices. For an in-depth derivation of all results in this section we refer
to~\cite{Buividovich:16:2}.  
The Jordan normal form of $\bA$ is given by 
\begin{equation}
  \label{eq:B_jdecomp}
  \mathcal{X}^{-1} \bA \mathcal{X} = \mathcal{J} \quad  \text{where} \quad \mathcal{J}  = \begin{pmatrix}
    J_1 & 0 & 0 & \dots & 0 \\ 0 & J_2 & \ddots & \ddots & \vdots \\ 0 & \ddots & \ddots &
\ddots & 0 \\ \vdots & \ddots & \ddots & \ddots & 0 \\ 0 & \dots & 0 & 0 & J_n
  \end{pmatrix} \quad \text{with} \quad J_i:=
  \begin{pmatrix} \lambda_i & 1 \\ 0 & \lambda_i
  \end{pmatrix}.
\end{equation}
The transformation matrices $\mathcal{X}$ and $\mathcal{X}^{-1}$ can be constructed
analytically in terms of the eigenvalues and (left and right) eigenvectors of the matrix
$A$ and their derivatives:

\begin{equation}
  \label{eq:transformation}
  \mathcal{X}^{-1}  = 
  \begin{pmatrix}
    \left(\LL{1}, {\dLL{1}} \right) \\
    {\pt \lambda}_1 \left(0, \LL{1}\right) \\
     \vdots  \\
     \left(\LL{n}, {\dLL{n}} \right)  \\
     {\pt \lambda}_n \left(0, \LL{n}\right)
  \end{pmatrix} \ \text{and} \
\mathcal{X}  = 
  \begin{pmatrix}
    \begin{pmatrix}
      \RR{1} \\
       0
    \end{pmatrix} \! \! \! \! &
    ,\frac{1}{{\pt \lambda}_1}
    \begin{pmatrix}
      {\dRR{1}} \\
      \RR{1}
    \end{pmatrix}  \! \! \! \!  &\! \! \!,\cdots,\! \! \! &
    \begin{pmatrix}
      \RR{n} \\
       0
    \end{pmatrix}   \! \! \! \! &
    ,  \frac{1}{{\pt \lambda}_n}
    \begin{pmatrix}
      {\dRR{n}} \\
      \RR{n}
    \end{pmatrix} \!
  \end{pmatrix}
\end{equation}
We can now use the known analytic expressions for the Jordan decomposition of $\bA$ to
devise a generalised deflation procedure. Instead of projecting out the eigenvectors
corresponding to the eigenvalues $\lambda_i$ close to zero, we project out the
subspace corresponding to the Jordan block $J_i$. For the sake of a compact notation let
$\ket{X_i}$ denote the columns of $\mathcal{X}$ and $\bra{\bar{X}_i}$ the rows of
$\mathcal{X}^{-1}$. Then $\sum_{i=1}^{2n} \ket{X_i}\bra{\bar{X}_i}=\mathbbm{1}$ and
we can define the projectors $\mathcal{P}_m = \sum_{i=1}^{m} \ket{X_i}\bra{\bar{X}_i}$  and
$\bar{\mathcal{P}}_m :=\mathbbm{1} - \sum_{i=1}^{m} \ket{X_i}\bra{\bar{X}_i}$, such
that every vector $\ket{\psi} \in \mathbb{C}^{2n \times 1}$ can be written as 
$\ket{\psi} = \mathcal{P}_m \ket{\psi}+\bar{\mathcal{P}}_m \ket{\psi}$. Likewise we can
write $ f(\bA)\ket{\psi} =
\mathcal{X}f(\mathcal{J}) \mathcal{X}^{-1} \mathcal{P}_{2l} \ket{\psi}
+f(\bA)\bar{\mathcal{P}}_{2l} \ket{\psi}$, where we use the projectors for the  $l$ eigenvalues
$\lambda_1, \cdots, \lambda_l$ closest to zero and equation~\eqref{eq:matrix_func}. The
function of the Jordan blocks is given by equation~\eqref{eq:Jordan_function} and by
exploiting the bi-orthogonality of  $\ket{X_i}$ and $\bra{\bar{X}_i}$ we finally get 
\begin{equation}
  \label{eq:final_res}
 \resizebox{.9\hsize}{!}{$ f(\bA)\ket{\psi} = \underbrace{\displaystyle
f(\bA)\bar{\mathcal{P}}_{\!{2l}}\ket{\psi}}_{\text{approximation}} +
\underbrace{\displaystyle{
\sum\limits_{i=1}^{l}\left[f(\lambda_i)\left(\ket{X_{\!{2i-1}}}\braket{\bar{X}_{\!{2i-1}}}{\psi}
+ \ket{X_{\!{2i}}}\braket{\bar{X}_{\!{2i}}}{\psi}\right)+ (\pt f(\lambda_i))
\ket{X_{\!{2i-1}}}\braket{\bar{X}_{\!{2i}}}{\psi}\right]}}_{\text{exact}}
$}.
\end{equation}
This looks almost like the deflation procedure for diagonalisable matrices, with the
difference that we have to use the vectors that make up the Jordan transformation matrices
instead of the eigenvectors.  Moreover because of the non-diagonal structure of the Jordan
blocks there are additional ``mixing terms'' proportional to $\pt f(\lambda_i)$. The sign
function is piece-wise constant and the derivative terms vanish, so that the generalised
deflation prescription for the sign function is given by  $\sgn(\bA)\ket{\psi} = \sgn(\bA)
\bar{\mathcal{P}}_{2l} \ket{\psi}  + \sum_{i=1}^{2l}\sgn(\lambda_i) \ket{X_{i}} 
\braket{\bar{X}_{i}}{\psi} $. Note that it is sufficient to know the first $l$ columns
(rows) of $\mathcal{X}$ ($\mathcal{X}^{-1}$) to perform the deflation. The required $l$ eigenvalues
and eigenvectors of $A$ can be computed with the Arnoldi method and in our
calculations we use the {\tt{ARPACK}} package for this task. The
calculation of the derivative of the eigenvectors is a bit more involved. The derivative
of a right eigenvector is given by $ \dRR{j} = \sum_{i \neq j} \frac{\RR{i}\LL{i} (\pt A)
\RR{j}}{\lambda_j - \lambda_i} $ and a similar equation holds for the derivative of the
left eigenvector. The summation over all eigenvectors can be avoided by using the fact
that  $\sum_{i \neq j}^n \frac{\RR{j}\bra{L_j}}{\lambda_i - \lambda_j}
= \sum_{i=1}^{(j-1)} \frac{\RR{j}\bra{L_j}}{\lambda_i - \lambda_j} +  \left(\lambda_i - A\right)^{-1}P_l^n $.

\begin{figure}[hb]
  \centering
\input{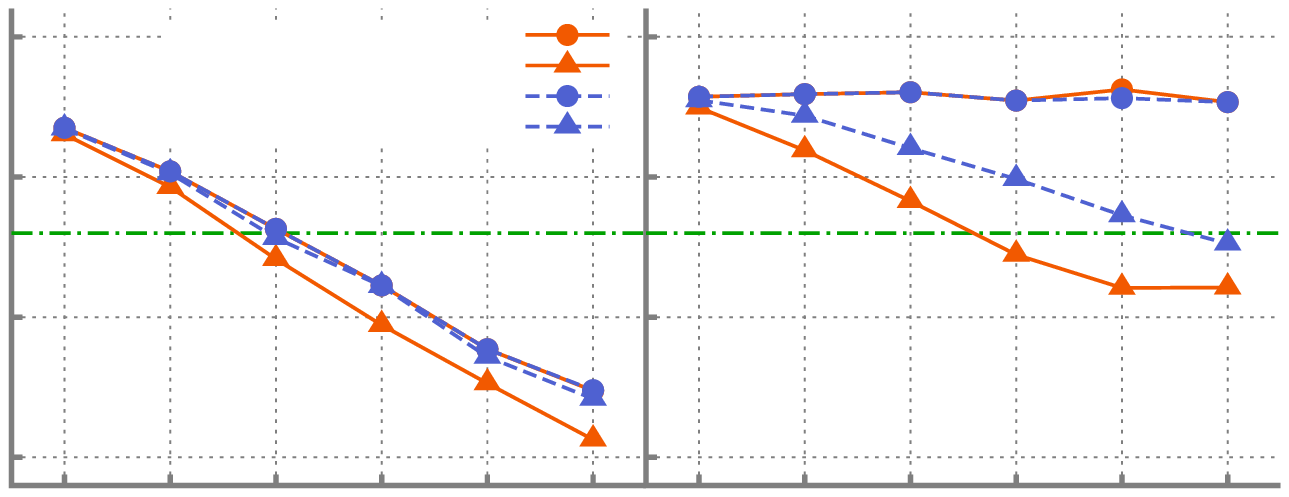}
  \caption{Results for the error of the nested TSL approximation (inner Krylov subspace
    size $100$) of the sign function and its derivative with and without deflation. The
    $40$ eigenvalues closest to zero are used to deflate the sign function. For the
    deflation of the derivative we use $6$ ($2$) eigenvalues for $V=14\times14^3$
    ($V=6\times18^3$).  On the left plot $V=6\times18^3$ and $\mu=0.230$ and on the right
    $V=14\times14^3$ and $\mu=0.300$. The dash-dotted green line marks an error of $10^{-7}$. }
  \label{fig:deriv_res}
\end{figure}

\section{Numerical Results}
\label{sec:res}
To test the efficiency our numerical differentiation method and the generalised deflation
prescription we compute derivatives of the overlap Dirac operator at finite chemical
potential. Our quenched $SU(3)$ gauge configurations are generated with the
tadpole-improved Lüscher--Weisz action~\cite{Luescher1985}. We use two different
parameter sets, a lattice volume $V=14\times 14^3$ with $\beta=8.1$ corresponding to a
temperature of $T=113 \ \MeV$ and $V=6\times 18^3$ with $\beta=8.45$ corresponding to
$T=346 \ \MeV$. The critical temperature for the deconfinement phase transition for the
Lüscher--Weisz action is around $300 \ \MeV$ and with our choice of parameters we have
configurations for both phases. The derivatives are taken with respect to an external
$U(1)$ lattice gauge field $\Theta_{\nu}(x)$. To approximate the matrix sign function we
use a nested version of the TSL~\cite{Bloch:11:2} with one level of nesting. The error of
the approximation is estimated using the fact that the square of the sign function is the
identity and we define the (relative) error of the sign function evaluation as $\epsilon_A
= \frac{\| \sgn(A)^2 \ket{\psi} - \ket{\psi} \|}{2 \ \| \ket{\psi} \|}$, where the factor
two in the denominator is used because we have to apply the TSL approximation twice to
compute the square of the sign function.  The error of the derivative is computed in the
same way, substituting $\bA$ for $A$ and the sparse vector $(0,\ket{\phi})^T$ for
$\ket{\psi}$.

In Figure~\ref{fig:deriv_res} we show the error of the function and the derivative as a
function of the (outer) Krylov subspace size. We find numerically that in the high
temperature phase the spectrum of the kernel operator already has a large gap around zero
and consequently the deflation does not lead to a big improvement. In the low temperature
phase on the other hand the gap in the spectrum is considerably smaller and we can clearly
see that the deflation greatly improves the efficiency of the TSL algorithm. Not only is
the error for the deflated approximation much smaller at a given Krylov subspace size, we
also observe that the improvement of the error is much bigger as the Krylov subspace size
is increased.

\begin{figure}[ht]
  \centering
  \input{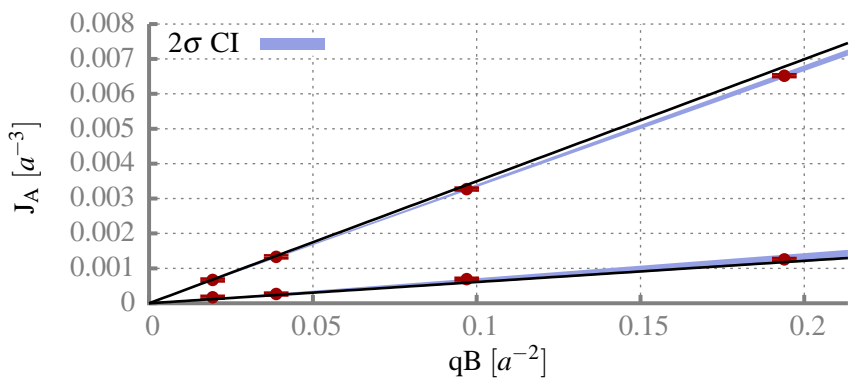}
  \caption{Results for the axial current in the chirally symmetric phase of QCD. The
    lattice size is $6\times18^3$ and $\beta=8.45$. The chemical potential is $\mu=0.040$
    and $\mu=0.230$ (steeper slope) in lattice units. Red dots mark the results for $J^A$
    and the blue bands the $2\sigma$ confidence intervals for $\scsc$. The black lines
    denote the result for non-interacting fermions in the continuum.}
  \label{fig:CI}
\end{figure}
\vspace{-10pt}
\section{Conclusion and Outlook}
\label{sec:conclusion}

In this work we have presented a method to take numerical derivatives of functions of general
complex matrices. We have shown how a generalised deflation prescription can be used to
improve the efficiency of the method and we have tested our method on quenched $SU(3)$
gauge configurations. Our main motivation to develop this method was the need to compute
conserved lattice currents for the overlap dirac operator in order to study anomalous
transport effects in dense QCD. In particular we are interested in the Chiral Separation
Effect (CSE)~\cite{Metlitski:05:1}, where an external magnetic field $B$ induces an axial
current parallel to the field in
a plasma of chiral fermions with finite chemical potential $\mu$: $J^A_i = \scsc B_i$. The
proportionality constant $\scsc$ is called the chiral separation conductivity and for free
fermions in the continuum it is given by $\scsc = qN_cN_f\mu/2\pi^2$. It is expected that
this relation also holds in the chirally symmetric phase of QCD, but there could be
corrections if chiral symmetry is spontaneously broken~\cite{Son:06:2}. On the lattice the
conserved axial current reads as $\langle J^A(x) \rangle = \Tr \left(\Dov^{-1}
\frac{\partial \Dov}{\partial \Theta_{x,\mu}}\gamma_5\right)$. To prove that our method
is well suited for practical calculations in Figure~\ref{fig:CI} we show first results for
the axial current and $\scsc$ in the chrially symmetric phase of QCD for two different
values of the chemical potential. The conserved axial current is not renormalised in
lattice calculations and we can directly compare our data with continuum results. Our
error bars are very small and we find a remarkably good agreement between our results and
the continuum value of $\scsc$. A more detailed analysis of the CSE in both phases of
QCD is work in progress and the results will be published elsewhere~\cite{Puhr:16:02}.

\end{document}